\documentclass[sigplan]{acmart}
\acmConference[CAIN 2024]{3rd International Conference on AI Engineering — Software Engineering for AI}{April 2024}{Lisbon, Portugal}
\usepackage{lscape}
\usepackage{graphicx}
\usepackage{soul}
\usepackage{enumitem}
\AtBeginDocument{%
  }

\copyrightyear{2024} 
\acmYear{2024} 
\setcopyright{acmlicensed}\acmConference[CAIN 2024]{Conference on AI Engineering Software Engineering for AI}{April 14--15, 2024}{Lisbon, Portugal}
\acmBooktitle{Conference on AI Engineering Software Engineering for AI (CAIN 2024), April 14--15, 2024, Lisbon, Portugal}
\acmDOI{10.1145/3644815.3644953}
\acmISBN{979-8-4007-0591-5/24/04}

\begin{document}

\title[Welcome Your New AI Teammate: On Safety Analysis by Leashing Large Language Models]{Welcome Your New AI Teammate: \\ On Safety Analysis by Leashing Large Language Models
%\thanks{Funded by Sweden's Innovation Agency, Diarienummer: 2021-02585.}
}

\author{Ali Nouri}
\orcid{0000-0002-9634-6094}
\affiliation{%
  \institution{Volvo Cars \& \\ Chalmers University of Technology}
  \streetaddress{P.O. Box 1212}
  \city{Gothenburg}
  \country{Sweden}
}
\email{ali.nouri@volvocars.com}

\author{Beatriz Cabrero-Daniel}
\affiliation{%
  \institution{University of Gothenburg, Department of Computer Science}
  \city{Gothenburg}
  \country{Sweden}}
\email{beatriz.cabrero-daniel@gu.se}

\author{Fredrik Törner}
\affiliation{%
  \institution{Volvo Cars,}
  \city{Gothenburg}
  \country{Sweden}
}
\email{fredrik.torner@volvocars.com}

\author{Håkan Sivencrona}
\affiliation{%
 \institution{Zenseact}
 \city{Gothenburg}
 \country{Sweden}}
\email{hakan.sivencrona@zenseact.com}

\author{Christian Berger}
\affiliation{%
  \institution{University of Gothenburg, Department of Computer Science}
   \city{Gothenburg}
 \country{Sweden}}
\email{christian.berger@gu.se}

\renewcommand{\shortauthors}{Nouri et al.}

\begin{abstract}
  DevOps is a necessity in many industries, including the development of Autonomous Vehicles. In those settings, there are iterative activities that reduce the speed of SafetyOps cycles. One of these activities is ``Hazard Analysis \& Risk Assessment'' (HARA), which is an essential step to start the safety requirements specification. As a potential approach to increase the speed of this step in SafetyOps, we have delved into the capabilities of Large Language Models (LLMs).
  Our objective is to systematically assess their potential for application in the field of safety engineering. To that end, we propose a framework to support a higher degree of automation of HARA with LLMs. Despite our endeavors to automate as much of the process as possible, expert review remains crucial to ensure the validity and correctness of the analysis results, with necessary modifications made accordingly.
%FT: we demostrate that LLMs can ... ?
\end{abstract}
%System Theoretic Process Analysis (STPA) which is an increasingly popular analysis method that emerges as a critical activity in autonomous vehicle engineering.


\received{November 2023}
\received[revised]{January 2024}
\received[accepted]{January 2024}

\maketitle

\section{Introduction}
% \textcolor{red}{Ready, please review, Anything else needed?}

The safety analysis of Autonomous Driving (AD) functions is crucial for engineers to identify hazardous events, assess their risks, and determine their root causes. Ensuring the safety of such functions often relies on iterative natural language (NL)-based activities, one of which is safety analysis. Artificial Intelligence (AI)-based tools capable of processing NL can be used to increase the efficiency and speed of these activities. One of the most promising tools is Large Language Model (LLM). 

Safety analysis is, however, not trivial as it consists of various activities such as identification of failure modes, and their effect in specific situations, which aim to mitigate or avoid unreasonable risks. Standards such as ISO 26262~\cite{ISO26262} or ISO 21448~\cite{sotif}, and regulations like UNECE R157 (ALKS)~\cite{ALKS} often propose or mandate activities such as Hazard Analysis Risk Assessment (HARA) and System Theoretic Process Analysis (STPA)~\cite{SEAASTPA}. These guides are used when specifying the safety requirements for mitigation or avoidance strategies. 

HARA is a well-known and usually required safety analysis for automotive functions, such as AD. The aim of this activity is to identify the hazardous events, categorize them, and to specify safety goals to prevent or mitigate them. Safety goals are top-level (i.e., vehicle-level) safety requirements that are then used in other safety activities. Together, these activities and intermediate results shed light on the safety of a function and the potential impact of a failure.

 Specific Operational Design Domains (ODD) are a starting point, where the AD function will be constrained to. The ODD begins to gradually expand once pieces of evidence are gathered and safety argumentation is implemented. In order to continuously update and improve the AD function and expand its ODD, the implementation of DevOps~\cite{google} is commonplace. Moreover, due to unknown hazardous scenarios~\cite{sotif}, DevOps serves as an approach to rapidly address their root causes as soon as they are identified. As it requires changes in function, system, and software, the corresponding safety activities shall also be addressed. This leads to the introduction of SafetyOps~\cite{18siddique2020safetyops}, representing the safety dimension within the DevOps framework. Depending on the updates in each iteration, SafetyOps may require modifications or the redoing of certain safety activities, including HARA.

However, the time needed for safety analysis and requirements engineering poses a significant challenge in achieving rapid SafetyOps~\cite{SEAAAliIndustrial}. This reinforces the need for new AI-based approaches that increase speed, reduce costs, and enable quicker responses to new events and incidents in the field. Rule-based approaches have significantly contributed to repetitive and well-established tasks, such as testing new software releases against legacy test cases. However, there are intellectual aspects in HARA such as identifying the consequences of a hazard in a specific scenario or specifying requirements, which cannot be automated by conventional tools.
%For instance all possible combination of scenario factors an be seen as an approach but, not all combinations make seance, and providing a rule which contain all logical relationships is almost impossible, and needs the involvement of a human to check feasibility of them. For instance the existence the existence of guardrail in urban road or a zebra crossing in a highway is quite rear or impossible. 
As mentioned before, LLMs are known for their strength in NL-based tasks, from academic writing~\cite{PromptEnginBio} to medical education~\cite{GPTMedicalExam}.
%On the other hand, they have also proven effective as tools for code generation, which is a more deterministic task.
LLMs could therefore be considered a potential approach to address the limitations of conventional tools in automating safety analysis. 

Communication with LLMs requires prompt engineering, whose domain-dependencies such as for supporting safety-related activities are currently under-explored~\cite{PromptEnginBio}. To the best of the authors' knowledge, no comprehensive and validated guides exist so far to guide practitioners in integrating LLMs into their safety analysis activities.
%Moreover, the strengths and weaknesses of utilized LLMs are still unclear and rapidly evolving.
Additionally, the use of these tools could soon be regulated by international regulations such as the AI Act~\cite{aiact}. Therefore, especially in safety-critical systems, the use of such tools must be thoroughly reported and addressed, with attention to confidence in the software tool process.

% AI Act versus this AI tool .... \textcolor{red}{@Bea, Do you know this? Can you write one sentec on it?}

\subsection{Background}

Item definition is the main input into HARA and includes a high-level description of the function and its interaction with the driver and environment.
Hazard identification is the first step in HARA. Then they are analyzed in various relevant scenarios to identify the consequences of hazardous events. If a hazardous event leads to any harm, the severity of this harm is classified with a value ranging from S0 to S3.
Finally, there arises a need to specify a safety goal once the hazardous event is assigned an Automotive Safety Integrity Level (ASIL) ranging from A to D.

\subsection{Research Goal \& Research Questions}
%\paragraph{Goals} 
The goal of this research is (i) to identify effective prompt patterns, and (ii) design a pipeline of prompts resulting in an LLM-based tool for performing HARA.
%and (iii) discuss with experts in the fields to what degree is the generated analysis acceptable according to the expectations at different stages.
Achieving this would enable the creation of an initial `version zero' of HARA. This version can then be expanded, modified, or used for brainstorming by development engineers, thereby accelerating the development of the valid version of HARA. This led to the following research question:

``What are the steps in a prompting pipeline to propagate the context needed to generate a HARA?''

%The participants were given the HARA, along with a review checklist and a comment sheet, to systematically review the HARA. 
In this paper, we present the designed pipeline of sub-tasks (i.e., prompts), serving as a guide for communicating with LLMs to achieve desired outcomes in analysis tasks such as HARA, and the lessons learnt during the design process.
%We also gathered a checklist and review protocol to quality assure the generated results, that was used to evaluate these results by engaging safety experts.

%The study is model-agnostic, treating the LLM as a black box, as it focuses on the prompts and their pipeline for LLM-based HARA. However, since the LLM itself and its training data significantly affect the results, if the study demonstrates the feasibility of LLM-based HARA for a novel automotive function. For this reason, we selected a novel automotive function (currently a concept) that, to the best of the authors' knowledge, has no external publications, and its HARA does not exist. Therefore, the leakage from the human-performed HARA is minimized to avoid indirectly guiding the LLM in performing HARA. 

%This study contributes to the safety domain by proposing prompts and a pipeline between them, which might reduce costs and increase the speed of performing safety analysis. Furthermore, it contributes to the prompt engineering field by validating some of the proposed patterns and introducing additional ones. The study is conducted transparently, providing explanations of the prompts and the pipeline. Subsequently, the output is verified by safety experts, and their comments are recorded.

%\paragraph{Structure of the Article} 
The rest of the paper is organised as follows:
Sec.~\ref{sec:relatedwork} presents related research.
%Sec.~\ref{sec:methodology} describes the methodology of this study.
Sec.~\ref{sec:Design_Cycles} addresses our research question by discussing the pipeline of prompts and prompt engineering.
%Sec.~\ref{sec:Engineering_Cycle} addresses RQ3 and reports the evaluation by experts and the results from the experts' quality assurance review. 
Finally, Sec.~\ref{sec:Threats} reports on threats to validity and Sec.~\ref{sec:conlusion} summarizes the findings and potential future studies.
%%%%%%%%%%%%%%%%%%%%%%%%%%%%%%%%%%%%%%%%%%
%%%%%%%%%%%%%%%%%%%%%%%%%%%%%%%%%%%%%%%%%%
%%%%%%%%%%%%%%%%%%%%%%%%%%%%%%%%%%%%%%%%%%
% attention
\section{Related Work}
\label{sec:relatedwork}
%explain how GPT works ....

%What is LLM: LLMs are probabilistic model of the next word based on the set of previous tokens.
LLMs are trained on a wide range of data across various domains and contexts, enabling them to generate text that is judged to be relevant for various tasks\cite{bommasani2022opportunities}. As prompt engineering plays an important role, there are some studies that propose specific prompt patterns such as \cite{promptPatterns2023}, which are discussed in detail in Sec.~\ref{subsec:Prompt Engineering}.
%There are some techniques for how to effectively use the conversation like Tree of Thoughts~\cite{ToT} which if used can improve the results. 

%The LLM is illustrated as \cite{PromptEnginBio} %a library which prompt engineering is the librarian
%Filtering could be done by using the known KPIs for each aspect in HARA and then for each cluster select the best row and select the best derived hazardous events which is part of a known technique named Tree of Thoughts.

However, some publications have reported the weaknesses of LLM such as hallucination~\cite{zheng2023chatgpt}, a phenomenon where the LLM generates text that does not align with reality.
This highlights the importance of identifying these weaknesses and their risks in each specific application.

To the best of the authors' knowledge, there is no systematic or peer-reviewed study that has assessed the capability of LLMs in performing safety analysis, specifically focusing on hazard analysis and risk assessment.
While there have been studies on using LLMs in STPA for Autonomous Emergency Braking (AEB)~\cite{AEBGPT}, and Hazard Analysis for a water tank system~\cite{diemert2023large}, these studies do not focus on the LLM's capability to perform the analysis autonomously; instead, they use it as a text generation assistant.
In both studies, the user continuously guided the LLM to perform the analysis and subsequently assessed the quality of the output.
This poses a threat to the validity of the study, as guiding the LLM can eventually lead to generating desirable text.
The second threat to validity arises from using mature functions, which the model has most likely encountered during training.
%This issue is discussed in more detail in Sec.~\ref{subsec:Problem Investigation}.

%How something logical can come out of such a concept (seems impossible), but lets see.

%The end2end ML is already examined in the industry (Nvidia), which it is claimed to be even safer since due to unbounded complexity of environment by having ML based approach instead of rule based the car is able to break even some rules (e.g., exiting the road) but keep the trajectory safe (e.g., avoiding collision with pedestrian). There are multiple issue with this approach such as violation of transparency and explainable requirement of AD (mobileye).

%Idea: but what if we let the ML based systems like LLMs to design a rule based approach and we used the human engineers to review the requirements, codes, and fix the issues if there is an issue there like non explainable parts of the software, or wrong code implementation.

%One idea is to ask LLM to directly write a code for the whole AD, but then the complexity of such a software would not let the human to well understand the software and then the bugs would not being captured. So we need the same abstarction levels as automotive has for human developers and same requirement hierarchy (our paper STPA). Moreover not all aspects of software are known like the hazards that the software might cause or the cause and effect relationship. Or the possible needs for the function limitations.

%%%%%%%%%%%%%%%%%%%%%%%%%%%%%%%%%%%%%%%%%%
%%%%%%%%%%%%%%%%%%%%%%%%%%%%%%%%%%%%%%%%%%
%%%%%%%%%%%%%%%%%%%%%%%%%%%%%%%%%%%%%%%%%%
% \begin{comment}
\section{Experiment Design} \label{sec:methodology}

% \begin{figure*}
%     \centering
%     \includegraphics[width=1\linewidth]{DesignEngineeringCycle.png}
%     \caption{The dotted blue arrows are showing the design cycle and the green arrows are showing the engineering cycle.}
%     \label{fig:DesignEngineeringcycle}
% \end{figure*}

In this study, we iteratively designed a pipeline for LLM-based HARA that comprises a set of carefully engineered prompts and their communication. We followed the recommendations of Hevner et al.\cite{DesignScienceHevner} and Roel J. Wieringa \cite{DesignScienceRoel} in this study. 
The study is designed to be model agnostic, although the experiment was conducted using GPT-4.0, developed by OpenAI, accessed through an API.

During the design cycles, the first author investigated the problem. Subsequently, the prompts and their pipeline were designed. The results were validated through internal discussions and review meetings with the other researchers in the team. 
During the initial phases, the study primarily focused on a feasibility study, as the researchers in the team, who had previously performed HARA, were aware of the task's difficulty and the intellectual capability required for it which was not expected from LLM.

For this experiment, we selected the fully unsupervised ``collision avoidance by evasive maneuver'' function (CAEM). The publicly available documentation for this novel function is less extensive than that for more mature functions like AEB, thereby reducing the likelihood that its HARA results are present in the training data of the LLM model under test.

We began by performing HARA for a CAEM function and gradually investigated the causes of issues, clarifying the question more thoroughly, which led to better results. As a result, the detail of each task increased to the point where decomposition of the steps was necessary.
Initially, we provided the scenarios as is common in industry. Once the model's capability was demonstrated, we advanced a step further and requested the scenario catalogue as well, providing only the guidelines for creating scenarios.

%``Design Science in Information Systems Research'' by Hevner et al.\cite{DesignScienceHevner} and ``Design Science Methodology for Information Systems and Software Engineering'' by Roel J. Wieringa
%Hevner et al.\cite{DesignScienceHevner} propose the following seven guidelines for design science, which are explained below for our study.

% \subsection{Design as an Artifact:}
% \label{subsec:Design as an Artifact:}
As it is shown in Fig.~\ref{fig:HARAPipeline}, the LLM-based HARA is designed to be automated without an engineer's intervention, where the input is the item definition (i.e., the function description), and the output is the HARA results. The HARA for the full function can be completed in less than a day due to automation, but it still requires review by human engineers.

\section{LLM-Based HARA}
\label{sec:Design_Cycles}

\begin{figure*}
    \centering
    \includegraphics[width=1\linewidth]{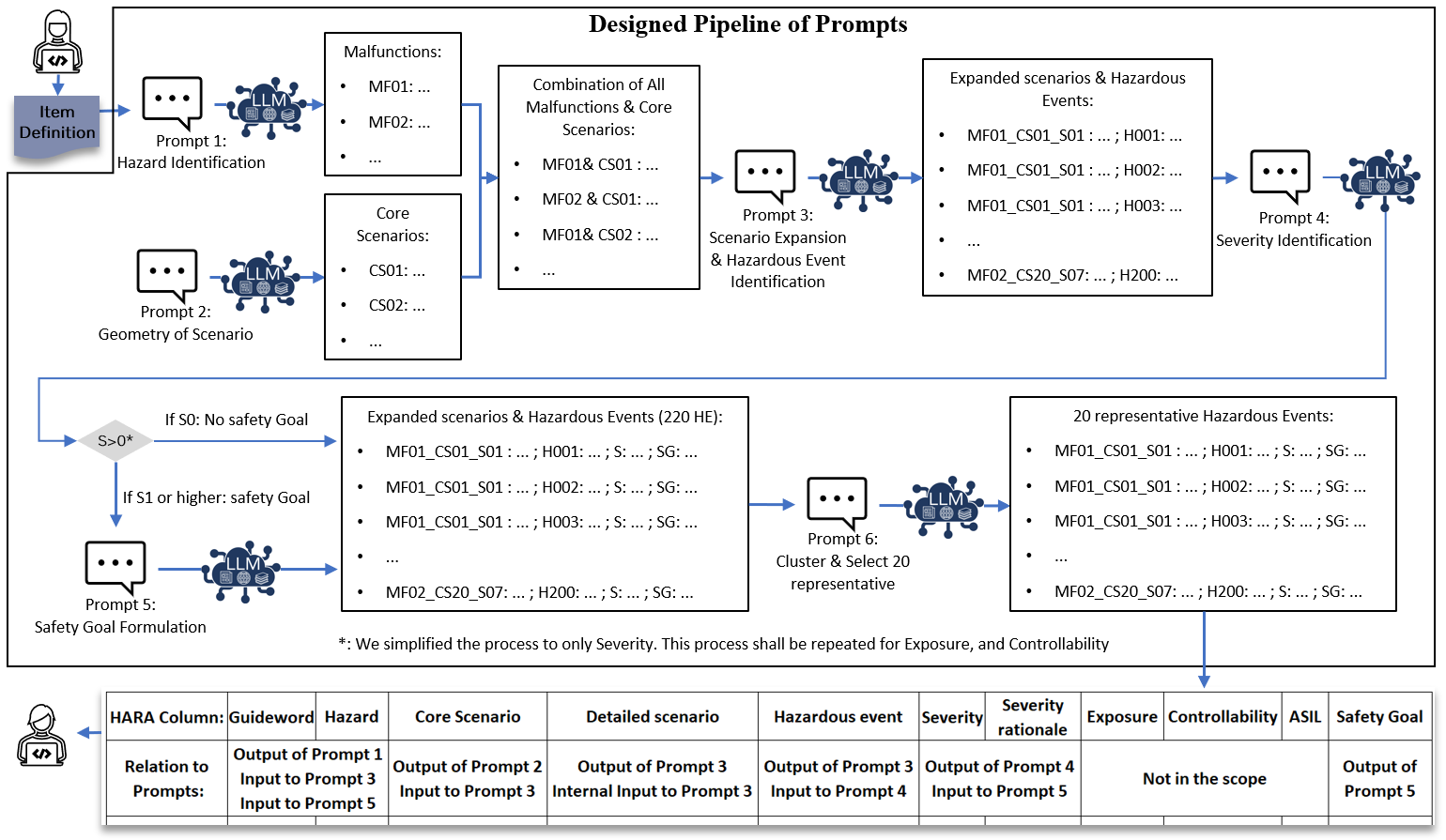}
    \caption{LLM-based HARA, utilizing a pipeline of subtasks, each managed through a specific prompt. The item definition is imported (top-left), and the HARA results are exported (bottom-right). In the second row of the HARA table, the relationship of each column to the prompts is summarized.}
    \label{fig:HARAPipeline}
\end{figure*}

Describing a task to an LLM requires specific skills, known as prompt engineering. The crucial skills for our application are detailed in Sec.~\ref{subsec:Prompt Engineering}.
Performing HARA necessitates diverse expertise for various subtasks, including scenario generation, identifying the consequences of hazards and their severity, as well as requirements engineering. Furthermore, each activity requires an extensive process description, which, in most cases, is potentially not publicly available and, as a consequence, is not included in the training data of LLMs.
Moreover, depending on the complexity of the function and its environment, this process could result in a large number of hazardous events and safety requirements.
Considering the token limitation constraint in LLMs, asking to perform HARA in the same prompt can lead to incomplete results.
So there is a need to break down HARA into subtasks and to design a pipeline to decompose and then integrate the results as explained in Sec.~\ref{subsec:HARA Pipeline} and presented in Fig.~\ref{fig:HARAPipeline}.

We aimed to design the prompts and the pipeline to receive the function description of any automotive function and to provide the HARA results without human intervention as shown in Fig.~\ref{fig:HARAPipeline}.
In the design of the prompts and pipeline, we considered the following restrictions (R1 to R4):

\textbf{R1: No input except the function description:}
Although scenario descriptions and malfunctions can often be extracted from company catalogs, this is not always the case, and it can significantly affect the quality of HARA. Therefore, we designed the pipeline to be as independent as possible, limiting inputs solely to the item definition.
For instance, we chose not to request scenarios as input, enabling the system to autonomously generate relevant scenarios.

\textbf{R2: No human intervention while HARA is on progress:} The analysis should be performed fully automated, without any intervention, allowing researchers or experts to review only the final results.
%It is also forbidden to modify the intermediate data between two prompts and the results of one sub-step shall be fed to the next sub-step without any change in the text. 
As the system's capability to produce the desired output might be compromised by human feedback or alterations to the intermediate data.
%of determining whether the LLM can independently perform safety analysis to an extent acceptable by the experts. 

\textbf{R3: No fine-tuning for specific function:} The prompts should only include the process to perform the analysis and not be tuned for any specific item definition, ensuring usability for any automotive safety-critical functions.

\textbf{R4: Automotive acceptable format:} The results of the analysis shall be presented in a format that is easily readable for reviewers. For instance, HARA should be extracted into a table, enabling reviewers to assess it systematically.

\subsection{Prompt Engineering}
\label{subsec:Prompt Engineering}

 To identify effective communication strategies with LLMs for performing the subtasks in HARA, we investigated prompt engineering.
This was done to craft the prompts and their pipeline effectively, aiming to improve the results in each design loop in comparison with the previous design loop.
After testing various prompt patterns in several iterations, the following patterns were found to be effective in achieving suitable results:

\textbf{P1:} Question Refinement Pattern\cite{promptPatterns2023} in forming the preliminary prompt: during the trial and error phase sometimes we used sequences of prompts to see the feasibility of getting the preferred results. When the preferred results were reached, the LLM was asked to provide the complete prompt needed to achieve the correct result in another conversation. 
Although the prompts were changed afterward and tuned to our needs with more details added, this approach accelerated the achievement of enhanced results.

\textbf{P2:} Question Refinement Pattern\cite{promptPatterns2023} in bug fixing of the finalized prompt: ``Why did you make this mistake?'' When LLM produced wrong inputs then the reason of the mistake was asked and normally the reason was contributing to the improvements of the new prompts and the same mistake was not seen again. The main groups of mistakes were due to wrong assumptions by the LLM.

\textbf{P3:} Breaking the tasks into sub-tasks:
A well detailed and concrete prompt, rather than a high-level one without any context, typically leads to better results.~\cite{promptPatterns2023}. This leads to longer prompts and given the limited number of tokens in each conversation, it is necessary to break down tasks into sub-tasks.
%but also to divide the requested large lists into subcategories of lists when needed.
%For example if you asked for the 1000 most effective scientists then it provide this answer and limit it to a number for its capability: ``Creating a comprehensive list of the 1000 most influential scientists and summarizing their achievements is a colossal task that cannot be effectively completed in this format. However, I can provide a shorter list ...''.

%for journal %\item[P4] Merge the sub-tasks: Although dividing a task is important but we need to be care-full, since some outputs are not relevant to the context and then producing useless outputs and later filtering them lead to the cost of the task. So if possible we asked for the two sub-tasks to be done at the same time. For example the expensive way is to ask for providing 100 scenarios for each core scenario and then filter the relevant ones for each malfunction. Instead we provided the malfunction and core scenario and asked for relevant detailed scenarios which lead to an incident. This not only reduce the cost but also lead to more coherent answer in both substeps.

\textbf{P4:} Reflection Pattern~\cite{promptPatterns2023}: 
%This pattern is as well for HARA that is done by human. 
It is necessary to provide a rationale for tasks such as severity classification, which assists reviewers in understanding the reasoning behind the selection of severities. This pattern is like reflection pattern~\cite{promptPatterns2023}, leading to better judgements of LLM.

There are other patterns that could improve the results; however, their use would conflict with the established rules and restrictions of our study.
For instance, the ``Cognitive Verifier Pattern''\cite{promptPatterns2023} encourages the LLM to ask follow-up questions for context clarification and to break down the question into sub-questions. However, this pattern necessitates user intervention during the pipeline's intermediate steps (a violation of R2) and demands additional inputs (a violation of R1).

%Although these patterns can be used for trial and is used in trial and error of the initial study but the aim of this study was to push the prompt engineering toward no human intervention till review stage. Our study was like a interview exam which the participant shall take the exam without any help and if we answer to the questions then the assessment is not accurate since there might be leakage of information in our answers. so this restriction help to see if LLM is capable of performing the HARA fully by itself. The other reason is that this part of the process would be used in a bigger picture which not always human intervention is possible. 

Each prompt is constructed with the following structure:

\begin{itemize}

\item Context: Part of the prompt where the context of the task is described, i.e., function description, definitions, and process.

\item Task: In this second part the task itself is described. Each task needs inputs from previous sub-steps which are indicated by placeholders and they are replaced automatically by the designed pipeline through the OpenAI API.
\item Template: At the end of the prompt the template of the response is provided to indicate in which format the output shall be produced which can be a table in Comma-separated values (CSV) format.
\end{itemize}

\subsection{HARA Pipeline}
\label{subsec:HARA Pipeline}

The pipeline of the prompts is designed as it is shown in Fig.~\ref{fig:HARAPipeline} and implemented in Python. In each step, the relevant inputs are taken from the previous steps which are recorded and used in the next step.
%in a Comma-separated values (CSV) file which a new column is added in each step gradually filled and is formed like a HARA template. 
%(see Supplementary Materials\footnote{The code and prompts (shall we keep the prompts as secret for the next paper? It has lost of process and other stuff) are accessible in, \url{TBD}}.)

\textbf{Input- Item Definition:} The pipeline starts from top-left corner of Fig.~\ref{fig:HARAPipeline} which item definition of the function is imported as the only input.

\textbf{Prompt 1- Hazard Identification:} The malfunction identification process is explained at the beginning of the context, followed by a placeholder for the item definition, which will be inserted. Then,``Your task: List the malfunctions of the function ...'' clarifies the task, and is followed by a table that serves as a template to be completed.

\textbf{Prompt 2- Geometry of Scenario:} The core scenarios are created by incorporating various aspects and categories of roads, such as the number of lanes, road shape, slope, and features like bridges. Then the task is to create 20 diverse road geometries followed by a template.

\textbf{Prompt 3- Scenario Expansion \& Hazardous Event Identification:} The malfunction and core scenarios are combined and then each combination will be placed in the task of this prompt. As the context the item definition is provided. 
The LLM is tasked with expanding the core scenario into more detailed scenarios that could potentially lead to harm, taking into account the provided malfunction.
The LLM is provided with a generic list of all possible objects and agents, along with their relative trajectories.

\textbf{Prompt 4- Severity Identification:} After outlining the process of severity identification, the LLM is asked to classify the severity and provide a rationale for it.

\textbf{Prompt 5- Safety Goal Formulation:} The item definition and the process of safety requirements specification are provided as a context.
Then the LLM is asked to specify safety goals for severities higher than zero.

\textbf{Prompt 6- Cluster\& Select 20 representative:} The hazardous events are grouped into four main categories, based on the two primary guide-words (i.e., Omission and Commission) and two main categories of severity (i.e., S0/S1 and S2/S3). Then the LLM was asked to find the five most representative hazardous event in each category.
%by proposing K-means \cite{aerospace7100143}.
%The number of the Hazardous Events (Rows of HARA table) depend highly on engineering judgement of safety experts which depend on the size of function, ODD, and other factors. So it would be benefical to implement a dimention reduction filter at the end of the pipeline. Moreover the incremental review process is recommended (ref our safetyOps paper) in SafetyOps process, which required a small portion of HARA to get the review comments faster, since although the HARA in the pipeline can be creating in less than a day but still the review process shall be done by human which delay the process if all HARA is provided to them. To satisfy this we implemeted a prompt to reduce the number of HARA rows from more than 200 to 20 representative ones. 

%. But in this stage we did not do that to better find the weaknesses which can be a good candidate for the last version of pipeline to remove the weak rows.

\textbf{Output- HARA Table:} The output is in the form of a table (in CSV format)\footnote{Providing  part of the LLM-based HARA without human intervention, \url{https://doi.org/10.5281/zenodo.10522786}}, as it is in industrial setup to be readable for the engineers, which is shown in the table on the bottom of Fig.~\ref{fig:HARAPipeline}.

\section{Threats to Validity}
\label{sec:Threats}

%No external validation, only iterations with internal reviews (however, knowledgeable!!), future work: design science extra cycles + case study.

As mentioned in Sec.~\ref{sec:methodology}, the evaluation has been conducted by researchers within the team to refine the pipeline and prompts.
However, further evaluation by independent experts external to the team, as part of the engineering cycle, is necessary and represents the next phase of this work.

\section{Conclusion \& Future Work}
\label{sec:conlusion}

% PAPER MESSAGE: "we designed a LLM-based HARA Tool which will provide the version zero of HARA and we report on strength and weaknesses, so the development teams in industry develop it further to a valid version"

% MAIN MESSAGE: This work tried to to create a LLM-based tool so automatically deliver the version zero/initial of HARA for being the starting point for development team to create the valid version
This study focuses on the use of LLMs, more precisely the GPT-4.0 model, in safety analysis activities such as HARA in order to accelerate the iterative loops in SafetyOps. The primary goal was to assess whether LLMs, through a systematic process using carefully engineered prompts, could deliver HARA results that meet the requirements of both industry standards and experts, even if human oversight is still required~\cite{aiact}. As a proof of concept, this study uses a novel AD function that was not part of the training data of GPT-4.0, to avoid trivial or biased safety goals. 

% MAIN MESSAGE: The automated HARA does not rely in additional inputs or human intervention.
Instead, all the contextual information was provided through the prompts that make up the HARA pipeline and a number of experiments were conducted to ensure that the information was correctly propagated through it. To achieve this, multiple design iterations were conducted to improve the HARA prompting pipeline. The designed HARA pipeline, described in Fig.~\ref{fig:HARAPipeline}, is able to provide an initial version of a safety analysis without the need for human intervention. % In this study, the human intervention and providing any input except item definition was not allowed to reduce the risk of biasing the LLM and to improve the results, thereby better capturing the weaknesses of the design. 
% MAIN MESSAGE: In order to check whether automated HARA is ok, we asked experts.
% MAIN MESSAGE: The experts said it is more than enough, however, they have comments.
While these results demonstrate the potential of LLMs in conducting HARA, they also highlights areas for improvement and the need for further development to fully meet industry standards and expert expectations. 

% \textcolor{brown}{Bea is here. She is defeated after many meetings. Bea will fix the next two paragraphs so everyone is motivated to keep reviewing the AI generations while making their safety analysis operations much faster and cheaper.}

% MAIN MESSAGE: However, remember that the HARA pipeline is meant to give a version zero of the analysis, not a full thing that you can truly trust.
Altogether, the results of this study show that GPT-4.0 is able to generate reasonable safety analysis that can be used as a ``version zero'' of the HARA. This version can then be expanded, modified, or used for brainstorming by development engineers, thereby accelerating the development of the valid version of HARA. This points outs to the need for posterior refinements by engineers, in line with the general recommendations for implementing human oversight strategies when using AI in safety contexts~\cite{aiact}.

Future work could, therefore, focus  on a more rigorous evaluation of the proposed pipeline by independent experts and on improvements of prompts and the pipeline to satisfy the experts' review comments. 

\section*{Acknowledgments}
Thanks to Sweden's Innovation Agency (Vinnova) for funding (Diarienummer: 2021-02585), and WASP, for supporting this work.

\section*{AI Ethics}
\label{sec:Ethical Considerations}
% \textcolor{red}{@Bea: Please review this}
%\textbf{Human participants in the study:} The the results of our proposal are evaluated by safety experts, so we followed the prescribed guideline for ethical principles by Strandberg \cite{swethic}. Each step of the study which human participants are involved was checked against the proposed checklist and relevant actions are done including anonymization, consent, and confidentiality. 
%\textbf{AI Ethics:}
The communication with LLMs and the evaluation of their capabilities is at the core of this work. It is therefore crucial to highlight the need to take AI ethics into consideration.
The nature of LLMs, and their rapid adoption and change rate, makes it difficult to completely explore their weaknesses and risks. Regulations such as the AI Act~\cite{aiact}, however, can guide their integration into systems or processes, as discussed in this paper.

Safety is also a source of concern for this LLM use case. For this reason, the function under analysis is novel and not yet commercialized. So the results of the analysis are only used in this study and not in production. The proposed pipeline and prompts are currently in the research phase and not used in any engineering of production related projects. The purpose, at this stage, is to conduct a feasibility study to assess whether further investigations are meaningful. %to use any of the results of this study in commercialized or real environment.
%until the technology is well mature and there are consunsue on usage of it in real environment from all aspects like safety, security, legal and so on.

\section*{Disclaimer}
The views and opinions expressed are those of the authors and do not necessarily reflect the official policy or position of Volvo Cars.
%Volvo Cars.

Please keep in mind that HARA is a complex process that involves understanding both the system and the environment which it is used in. LLMs like GPT-4.0 are able to assist with this process, as discussed in this paper, but they cannot fully understand or model the system in the way a human expert would. Therefore, they shall be used cautiously and their output must always be reviewed and validated by human experts.

%%b%%\textcolor{red}{Fix references}
%%%%%%%%%%%%%%%%%%%%%%%%%%%%%%%%%%%%%%%%%%%%%%%%%%%%%%%%%%%%%%%%%%%%%%%%%%%%%%%%%%%%%%%%%%%%%%%%%%%%%%%%%%%%%%%%%%%%%%%%%%%%%%%%%%
%%%%%%%%%%%%%%%%%%%%%%%%%%%%%%%%%%%%%%%%%%%%%%%%%%%%%%%%%%%%%%%%%%%%%%%%%%%%%%%%%%%%%%%%%%%%%%%%%%%%%%%%%%%%%%%%%%%%%%%%%%%%%%%%%%

%\bibliography{HARALLM} 

\begin{thebibliography}{10}

\bibitem{ISO26262}
``{ISO 26262:2018 (all parts), Road Vehicles — Functional Safety},'' standard, International Organization for Standardization, 2018.

\bibitem{sotif}
``{ISO 21448:2022, Road Vehicles — Safety of the Intended Functionality},'' standard, International Organization for Standardization, 2022.

\bibitem{ALKS}
{UNECE}, ``{UN Regulation No. 157 - Automated Lane Keeping Systems (ALKS)},'' 2021.

\bibitem{SEAASTPA}
A.~Nouri, C.~Berger, and F.~Törner, ``{On STPA for Distributed Development of Safe Autonomous Driving: An Interview Study},'' in {\em {Proceedings of the 49th EUROMICRO Conference on Software Engineering and Advanced Applications (SEAA)}}, (DURRES, ALBANIA), Sep. 2023.

\bibitem{google}
Google, ``{What is DevOps?},'' 2022.

\bibitem{18siddique2020safetyops}
U.~Siddique, ``{SafetyOps},'' {\em arXiv preprint arXiv:2008.04461}, 2020.

\bibitem{SEAAAliIndustrial}
A.~Nouri, C.~Berger, and F.~Törner, ``{An Industrial Experience Report about Challenges from Continuous Monitoring, Improvement, and Deployment for Autonomous Driving Features},'' in {\em Euromicro Conference on Software Engineering and Advanced Applications}, pp.~358--365, 2022.

\bibitem{PromptEnginBio}
L.~Giray, ``Prompt engineering with chatgpt: A guide for academic writers,'' {\em Annals of Biomedical Engineering (2023)}.

\bibitem{GPTMedicalExam}
A.~Gilson, C.~Safranek, T.~Huang, V.~Socrates, L.~Chi, R.~Taylor, and D.~Chartash, ``How well does chatgpt do when taking the medical licensing exams? the implications of large language models for medical education and knowledge assessment,'' 12 2022.

\bibitem{aiact}
{European Commission, Directorate-General for Communications Networks, Content and Technology}, {\em EUR-Lex - 52021PC0206 - EN - EUR-Lex}.
\newblock CNECT, 2021.

\bibitem{bommasani2022opportunities}
R.~Bommasani {\em et~al.}, ``On the opportunities and risks of foundation models,'' 2022.

\bibitem{promptPatterns2023}
J.~White, Q.~Fu, S.~Hays, M.~Sandborn, C.~Olea, H.~Gilbert, A.~Elnashar, J.~Spencer-Smith, and D.~C. Schmidt, ``A prompt pattern catalog to enhance prompt engineering with chatgpt,'' 2023.

\bibitem{zheng2023chatgpt}
O.~Zheng, M.~Abdel-Aty, D.~Wang, Z.~Wang, and S.~Ding, ``Chatgpt is on the horizon: Could a large language model be suitable for intelligent traffic safety research and applications?,'' 2023.

\bibitem{AEBGPT}
Y.~Qi, X.~Zhao, S.~Khastgir, and X.~Huang, ``Safety analysis in the era of large language models: A case study of stpa using chatgpt,'' 2023.

\bibitem{diemert2023large}
S.~Diemert and J.~H. Weber, ``Can large language models assist in hazard analysis?,'' 2023.

\bibitem{DesignScienceHevner}
A.~R. Hevner, S.~T. March, J.~Park, and S.~Ram, ``Design science in information systems research,'' {\em MIS Quarterly}, vol.~28, no.~1, pp.~75--105, 2004.

\bibitem{DesignScienceRoel}
R.~Wieringa, {\em Design Science Methodology for Information Systems and Software Engineering}.
\newblock 01 2014.

\end{thebibliography}
%\bibliographystyle{ieeetr}
%\footnote{The 20 representative HARA items are accessible in, \url{}}.

\end{document}